\documentclass[12pt]{iopart}
\usepackage{epsfig}

\usepackage{rotating}

\usepackage{upgreek}
\newcommand{\uHz}{\ensuremath\upmu{\mbox{Hz}}}

\begin{document}

\article[LISA Science Results/Data Disturbances]{}
{LISA Science Results in the Presence of Data Disturbances}

\author{Scott E Pollack}
\address{JILA, University of Colorado, Boulder, CO  80309-0440}
\ead{pollacks@jilau1.colorado.edu}

\begin{abstract}
Each spacecraft in the Laser Interferometer Space Antenna (LISA)
houses a proof mass which follows a geodesic through spacetime.
Disturbances which change the proof mass position, momentum,
and/or acceleration will appear in the LISA data stream as 
additive quadratic functions.  These {\it data disturbances}
inhibit signal extraction and must be removed.  In this paper
we discuss the identification and fitting of monochromatic
signals in the data set in the presence of data disturbances.
We also present a preliminary analysis of the extent of 
science result limitations 
with respect to 
the frequency of data disturbances.
\end{abstract}

\submitto{\CQG}

\section{Introduction}\label{intro}

The Laser Interferometer Space Antenna (LISA) will detect and study in detail
gravitational waves from cosmological sources \cite{PPA}.  
Probably the most interesting sources that LISA will detect are 
binaries containing massive black holes.  
Merging galactic structures may contain intermediate mass or 
more massive black hole binary systems.  As the
binary inspirals it will radiate gravitational waves at low frequencies, 
typically below 10~mHz for much of the inspiral phase.  Systems involving
supermassive black holes will radiate at still lower frequencies,
down into the $\uHz$ frequency region.  Gravitational wave signals such as these
will help to answer questions about galaxy formation and massive
black hole production.

The LISA constellation will orbit the Sun some fifty million kilometers
away from the Earth.  Although far from 
disturbances due to the Earth-Moon system,
the LISA environment is not completely benign.  
There are two environmental considerations
for LISA that will be especially important to us: cosmic rays and
large solar flares.  The probability of a micro-meteorite encounter
with one of the spacecraft is very small, and
it is unlikely that collisions will be significant compared to the
other two events.

Cosmic rays from the galaxy and Sun bombard each LISA spacecraft continuously.  
Cosmic rays with energies above about 100~MeV will be able to travel through 
the spacecraft 
shielding and deposit some of their energy and charge in the enclosed 
proof mass.  
Charging of the proof mass
is a problem due to the resulting electrical and magnetic forces 
on the proof mass.
As a gravitational wave antenna, the sensitivity of LISA is strongly
dependent on each proof mass following a geodesic.  To counteract
charging of the proof mass, it is planned to continuously or
periodically discharge the proof mass using a discharging lamp which shines
ultraviolet light onto the proof mass and/or its housing, 
ejecting photoelectrons.

Throughout the 22 year solar cycle there are times when the Sun is 
particularly active.  At these times, the Sun will occasionally
produce large solar flares.  These solar flares produce a large
increase in solar 
cosmic rays.  
During every large solar flare event, and possibly during discharging,
the proof mass will get a momentum kick.  During the event the
proof mass will be slowly pushed in some direction and viable
data may not be obtained.  At the end of the event the 
proof mass will have acquired a new position,
and will have a new velocity.  

In addition to the environmental disturbances, the
telecommunication antenna (which communicates with Earth)
periodically (perhaps as often as once per week) 
will be rotated by 0.1~radians or more.
This movement temporarily will disturb the spacecraft and 
the disturbance reduction system, causing the 
proof masses to have changes in their relative positions and velocities.
After the antenna has been rotated there will 
be a new gravitational field around the proof mass
which will cause the proof mass to have a new acceleration.
This is because the mass distribution of the antenna will not be 
known well enough to calculate the change in the gravitational
acceleration accurately.

The changes in the positions, velocities, and accelerations of the proof
masses will appear in the LISA data stream.
The new positions will change the number of laser light wavelengths
between each proof mass and therefore will be an offset 
in the LISA data stream.  A new velocity of the proof mass
will appear as a slow linear ramp of the data, and an
acceleration will appear as a quadratic term.  

These complications of the LISA data stream will hinder data analysis
if not removed.  Since we will have no other way to know the changes
of the positions, velocities, and accelerations of the proof masses
to sufficient accuracy, we must fit for these parameters in the data.  
In this paper we demonstrate a fairly simple
method for locating a gravitational wave signal in the LISA data
stream in the presence of a data disturbance.  
In \S\ref{LISA} we describe the construction of our simulated LISA data stream;
in \S\ref{search} we describe our search algorithm, and the methods used
for locating the signals and disturbances; in \S\ref{science} we apply our
search algorithm to our simulated LISA data stream; in \S\ref{limitations}
we analyze the affect of 
multiple disturbances on the science results.

\section{Simulated LISA Data Stream}\label{LISA}

Our simulated LISA data stream consists of noise, a monochromatic signal, and a
disturbance.  This section describes each of the three components in detail.

\subsection{Noise Spectrum Realization}

We use a suggestion for the LISA sensitivity curve described in \cite{Bender} 
which extends down to 3~$\uHz$.  We further extend this curve to 0.3~$\uHz$.
Our noise realization also includes an approximation for the binary 
confusion noise
present in our galaxy \cite{Hils}.  Above 3~mHz the 
sensitivity is determined mainly by white noise in measuring 
the distances between the proof masses, and is
dominated by shot noise in the LISA lasers.  The spectrum from 3~mHz down
to 0.1~mHz is dominated by the binary confusion noise.  This part of the
spectrum rises steeply at first and then roughly as $1/f$ from 1.6 to 0.1~mHz.  
Below 0.1~mHz the spectrum is dominated
by acceleration noise, with a power law exponent of $-5/2$.  
Thermal and proof mass
charge fluctuations increase the acceleration noise at frequencies 
of 10 to $3~\uHz$
with another power of $\sqrt{f}$.  It is assumed here that the spectrum
continues to rise as $1/f^3$ down to 0.3~$\uHz$.

We create a realization of the LISA sensitivity curve in frequency space
by generating gaussian deviates to multiply each amplitude.  
Inverse Fourier transforming the spectral amplitude gives us a times series.
Our time series contains $2^{15}$ points sampled at 100~second intervals, 
making the observation time just under 38~days.  Throughout this paper
$t_i$ is the time stamp
at measurement $i$, and $y_i$ is the simulated LISA data stream with
disturbance and signal at $t_i$.

The bulk of our analysis used one realization of the LISA data stream.
To gain confidence in our search algorithm we produced 35 different
realizations of the LISA data stream.  If not otherwise mentioned,
all results refer to the first realization.  

\subsection{Signal and Disturbance}

For this analysis we have chosen to look at single monochromatic sinusoids 
with constant amplitude.  
The actual LISA signals may be considerably more complex.  For instance,
there may be a large number of separable signals below $100~\uHz$, and many
may change their frequencies considerably in one month.  However, information
for the single signal case is expected to give some indication of
what is likely to happen in more complicated cases.

We analyze two single signal cases.  The first signal has a frequency
of $3~\uHz$, which is near the ``limit'' of the LISA sensitivity.  
This signal contains
nearly 10 complete cycles within our 38~day period.  
In view of the low frequency,
this signal should give us strong indications of limitations on the 
frequency of data disturbances.

The second case has a signal at $100~\uHz$.  As the LISA constellation
orbits the Sun gravitational wave signals will be modulated in
amplitude and frequency.  
However, the wavelength corresponding to $100~\uHz$ is 20~AU, so the
signal will have minimal spread in frequency space.  Therefore
we are justified in approximating a signal at this 
frequency as monochromatic.

Our signals are characterized by an amplitude, a frequency, 
and a phase.  We represent the signal symbolically as a 
function $h (t_i; A, f, \phi)$.
We define a disturbance as an interruption in the data stream where
no data is present, and where at the end of the interruption the
proof mass may have a new position, velocity, and acceleration.
The interruption has a duration of length $L$.  The disturbance
is characterized by a starting time, an ending time, and three
parameters $[p, u, a]$ which are the additional position, velocity,
and acceleration of the proof mass.  We represent the disturbance
symbolically as a function $g (t_i; p, u, a)$.

\begin{figure}[!hb]
\centering
\epsfig{file=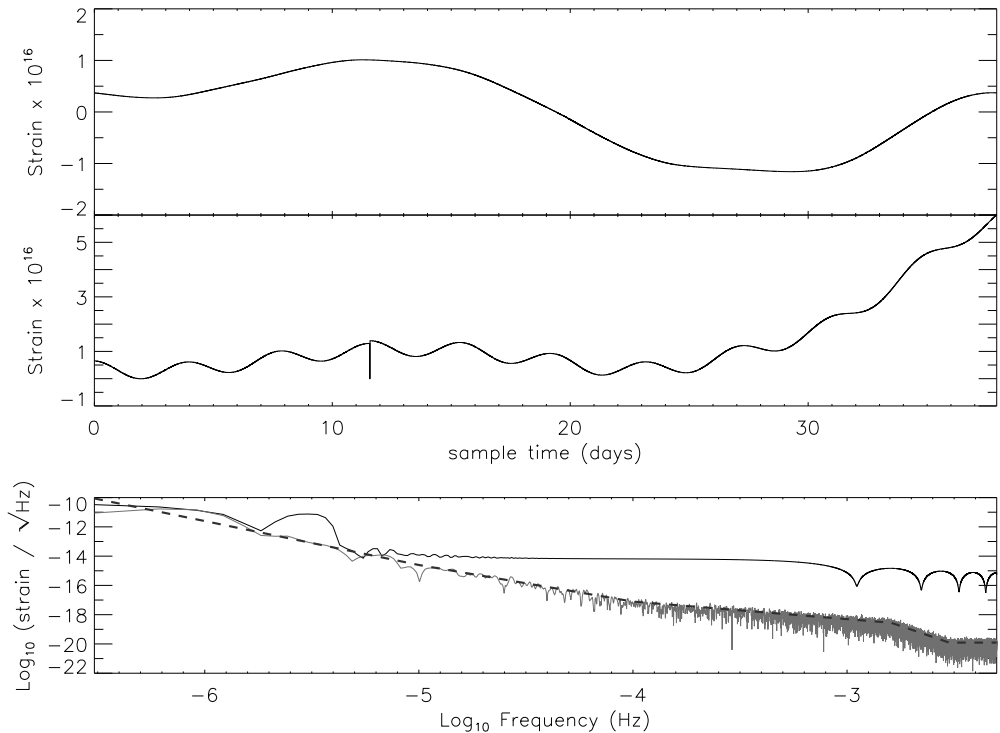, width=4in, angle=90}
\caption{\label{realization} The top frame
depicts one realization of the LISA noise spectrum
in the time domain.  The center frame adds to the noise
a monochromatic sinusoid with $(A,f,\phi) = (2\times10^{-17}, 3\uHz, 0^\circ)$,
and a disturbance with $[p, u, a] = [10^{-17}, 10^{-23}, 10^{-28}]$ located
at 11.6~days with an interruption length of 17 minutes.
The bottom frame shows the spectral amplitudes of the
top two frames, the dotted line is our LISA sensitivity curve.
}
\end{figure}

Figure~\ref{realization} shows one realization of the LISA data stream
with a monochromatic signal at $3~\uHz$ and a disturbance located
at 11.6~days after the beginning of the data set.

\section{Search Algorithm}\label{search}

We use an iterative approach to finding the signal and disturbance parameters.
Our algorithm utilizes both the downhill simplex method of Nelder and Mead
\cite{NelderMead, NRC} and the directional downhill decent method of 
Marquardt \cite{NRC}.  Given an initial collection of possible solutions
to the search problem (generated randomly) we use the Nelder and Mead
method to quickly find a local minimum of the problem.  

As explained in \cite{Lagarias} the Nelder and Mead method is very useful
for producing a rapid initial decrease in estimator values in 
low parameter dimensions.
But in higher dimensions the Nelder and Mead method fails to find
minimized estimator values.  However, the Nelder and Mead method can be forced
to continue minimizing the estimator with prodding.  Some of our prodding
is randomly restarting our simplex centered about the previous solution.
Other prodding involves modifying the estimator. 

Once the Nelder and Mead method ceases to refine the solution we 
turn to the directional downhill decent method of Marquardt.  
This search method uses derivative information to minimize the estimator.
Finishing with the Marquardt method guarantees a minimized estimator, typically
with a value much less than that provided by the Nelder and Mead method.

\subsection{Evaluation Criteria}

The search algorithm is dependent on an evaluation criterion which determines
how likely a solution is to be the answer.  
Sine we have correlated noise, we cannot determine an uncertainty associated
with each measurement point in the time domain.  But we do have some 
knowledge of the noise: we know its frequency spectrum.  Therefore
we can construct a $\chi^2$ in the frequency domain with a diagonal
correlation matrix:
$\chi^2 = \sum_{i = 0}^{N - 1} \left[r (f_i) / \sigma (f_i)\right]^2$,
\noindent where $r (f_i)$ is the Fourier transform of $y_i - h (t_i) - g (t_i)$,
and $\sigma (f_i)$ is our LISA sensitivity curve.

There are two modifications we make to our estimator during our search.
The first is to window the data before taking the Fourier transform.  
The discrete Fourier transform of a sinusoid is a delta function in
frequency space with large wings dropping off as $1/f$.  Using a modified
Hanning window we suppress the wings of the function, but we also
spread the peak of the function out to cover more frequency bins.
This modification helps the Nelder and Mead method to locate the frequency
and amplitude of the sinusoid in our data stream.

The second modification we make to our estimator is to weight the residuals.
After the Nelder and Mead has found decent parameter values we no longer
window the data.  Our search algorithm can get away with minimizing the $\chi^2$
by lowering the frequency of the sine wave, which has the effect
of reducing the transform wings at high frequencies.  
Since there are many more points at high frequencies than there
are at low frequencies the value of $\chi^2$ will be reduced, but there
will be extra residuals at low frequencies, since the sine wave frequency
is not being fit appropriately.  To force our algorithm to fit the sine
wave frequency we weight the residuals at low frequencies more than the
residuals at high frequencies.  We justify this weighting by the fact
that we know we are looking for a low frequency sine wave, so we 
preferentially want to reduce the $\chi^2$ contribution from the low
frequency residuals.

In $\chi^2$--tests a reduced $\chi^2$ of 1 indicates a good fit.  We
have 32768-point data sets, with 6 parameters to be fit, which
yields 32762 degrees of freedom.  Any value of $\chi^2$ around this value
is considered a good fit.  But how good is good?  We would like
some supercriterion to determine how well the search algorithm is
actually performing.
Our supercriterion compares the algorithm's
fit for the signal, in the presence of both noise and a disturbance, to
the algorithm's fit for the signal in the presence of noise only:
$\Delta^2 = \sum [h(t; A, f, \phi) - h(t; A, f, \phi)_\mathrm{nodis} ]^2$.
In this way we can judge how well different search criteria and
search algorithm's remove the disturbance from different data streams.

\subsection{Disturbance Identification and Removal}\label{nosig}

We know the location and duration of each disturbance with some certainty.
Our search algorithm is fed the locations and durations of each disturbance
and is then asked to remove those disturbances as best as possible.
The bulk of our analysis had only one disturbance present in the data set.
The disturbance function is a quadratic.  It easily can be shown that the
accuracy to which you can fit a quadratic is dependent on the number of
points available.  Therefore the location of the disturbance within our
38-day data set should affect our ability to accurately identify the disturbance
parameters.  

We examined three disturbance locations in detail.  For the
rest of this article we refer to the three disturbances as $g1$, $g2$,
and $g3$.  
$g1$ has a disturbance start time of 84,000 seconds (0.97 days)
after the beginning of the time series.
$g2$ has a disturbance start time of 500,000 seconds (5.8 days)
after the beginning of the time series.
$g3$ has a disturbance start time of 1,000,000 seconds (11.6 days)
after the beginning of the time series.
Each of these disturbances had the following parameters 
$[p, u, a]_\mathrm{input} = [10^{-17}, 10^{-23}, 10^{-28}]$,
which are referred to as the input disturbance parameters.  These values
correspond to a position change of about 26~nm, a velocity change of about 
26~fm/s, and an acceleration change of about 0.26~attometer/s$^2$
of the proof mass.  The length of the data interruption is 
ten data points, i.e., 1000 seconds, or about 17 minutes.

\begin{sidewaysfigure}[p]
\centering
\epsfig{file=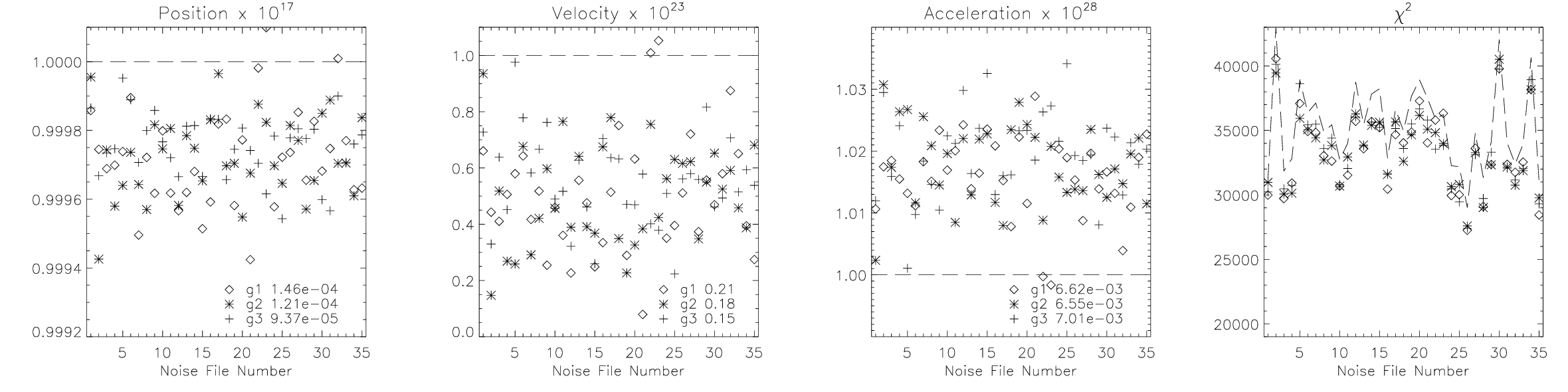, height=9in, width=2.in, angle=90}
\caption{\label{gapOnlyFig} Search algorithm results for the three disturbance
locations examined, in the 35 realizations of the LISA data stream.  The 
dashed lines represent the input parameters.  In the case of $\chi^2$, 
the dashed line is the value of $\chi^2$ in the 
original LISA data stream realization.
From the collection of 35 values for each parameter we compute the standard
deviation.  We can take this number as an indicator of the error in our derived
quantity.  This number is printed at the bottom of each plot, for each 
disturbance location.}
\epsfig{file=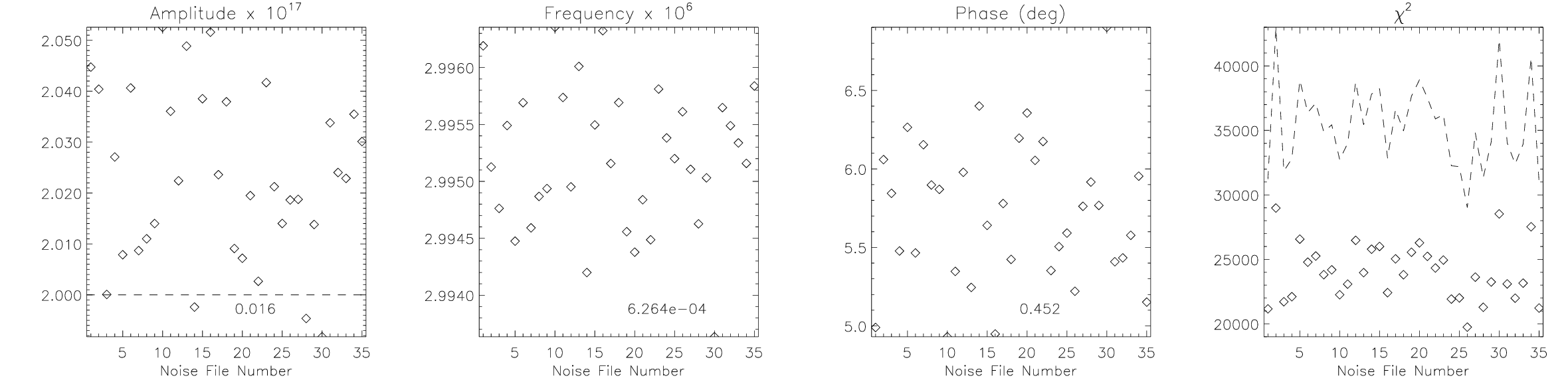, height=9in, width=2.in, angle=90}
\caption{\label{snr} Search algorithm results for 
the 35 realizations of the LISA data stream
in the presence of a SNR=5 signal only.  The 
dashed lines represent the input parameters.  In the case of $\chi^2$, 
the dashed line is the value of $\chi^2$ in 
the original LISA data stream realization.
The standard deviation of each collection is printed at 
the bottom of each plot.}
\end{sidewaysfigure}

Table~\ref{gapOnly} shows the results of our search algorithm when no signal
is present.  The determination of the position is very good, and the derived
velocity and acceleration make a smooth residual curve throughout the 
interruption region.  
Note that the algorithm has found a lower $\chi^2$ than obtained
by using the input parameters.  Figure~\ref{gapOnlyFig} contains our results
for all 35 noise realizations.  We can use the standard deviation of the derived
parameters throughout the 35 noise realizations as an estimate of our error
in any one derived parameter.  The errors are roughly 
$[0.01\%, 18\%, 0.67\%]$.
Therefore we can conclude that our algorithm finds confident values
for the position and acceleration, but the velocity may not be accurate.
Note that in figure~\ref{gapOnlyFig} the acceleration is systematically
higher than the input value, and that the velocity (by correlation is therefore)
lower than the input value.  This slight shift in the acceleration and velocity
has allowed our algorithm to find a lower $\chi^2$ value.
The fact that we have found ``incorrect'' values for the disturbance
parameters should not deter
us since the uncertainties should not depend on the assumed values, 
and since our main goal is to see if and how this affects the science results.

\begin{table}
\caption[Disturbance and Noise Only]{\label{gapOnly}
Search results for the three disturbance locations examined with
no signal present.  In all three cases, the search algorithm has minimized
$\chi^2$ below that which the input parameters give.
$p^* = (p - 10^{-17}) \times 10^{20}$, 
$u$ is in multiples of $10^{-23}$, $a$ is in multiples of $10^{-28}$.}
\begin{indented}\lineup
\item[]\begin{tabular}{@{}|c||rrr||c|}
\hline
 & \multicolumn{1}{c}{$p^*$} & \multicolumn{1}{c}{\0$u$} & \multicolumn{1}{c||}{$a$} & 
   $\chi^2$ \\
\hline \hline
\input{gapOnly.out}
\hline
\end{tabular}\end{indented}\end{table}

\subsection{Signal Identification and Removal}

For our realization of the LISA data stream, an rms strain
amplitude of $2 \times 10^{-17}$ has a signal-to-noise 
ratio (SNR) of 5 for a 3~$\uHz$ signal.
We define the SNR using the frequency spectrum of the data set.  The
SNR is the amplitude of the sine wave divided by the root mean squared 
noise level at the frequency of the sine wave.
We injected a signal with this amplitude, frequency, and a phase of zero
degrees into our data stream, so that
$(A, f, \phi)_\mathrm{input}~=~(
2\times10^{-17}, 3\uHz, 0^\circ)$.  
Without a disturbance in the data stream,
our search algorithm found these signal parameters
$(A, f, \phi)_\mathrm{nodis}~=~(
2.045\times10^{-17}, 2.996191\uHz, 4.990^\circ)$
with $\chi_\mathrm{nodis}^2~=~21164$.  
These will be the parameters used
in our calculation of our supercriterion $\Delta^2$.  
Using the input signal parameters we found 
$\chi^2_\mathrm{input}~=~31219$.  
Taking the sum of the squares of the difference between the solution and 
the input signal gives a value of $0.437\times10^{-31}$.  
The correlated noise in the data stream ``pulls''
the solution away from the input signal parameters.
In effect, our search algorithm
has been able to lower $\chi^2$ by fitting some of the noise.

Figure~\ref{snr} displays the results from our search algorithm for all 35 noise
realizations in the case of a signal only, with no disturbance.  
This data shows how well our algorithm can find the $3~\uHz$ signal out
of the noisy data.  We located the amplitude to within 1\%, the frequency
is consistently low by about 2\%, and the phase is correspondingly high
by about 6 degrees.  The systematic error in our derived frequency is a 
result of the correlated noise rising as $1/f^3$ at lower frequencies.
Our algorithm is able to achieve a lower $\chi^2$ by fitting a slightly
lower frequency sine wave.  The sine wave frequency is anti-correlated
with the phase, so the phase is correspondingly high when the frequency
is low.  This systematic error should not deter us in this investigation
because we wish to find the limitations that disturbances add to our
identification of the science signal, not necessarily how well we
can fit the sinusoid individually.
\section{Effects on Science}\label{science}

We now examine the full LISA data stream with a monochromatic signal and
a disturbance.

\subsection{Lower Frequency Signal}\label{lowfres}

A strong low frequency signal can alter the determined disturbance parameters.
This is why we utilize random restarts of our simplex to prevent becoming
caught in local minima.
In this way we can better determine all six parameters of interest.  
Table~\ref{ge3} contains our search results for the signal 
and disturbance parameters for the
three disturbance locations examined.  Our search algorithm was again able
to minimize $\chi^2$ below that which the input parameters yield.  

\begin{sidewaysfigure}[p]
\centering
\epsfig{file=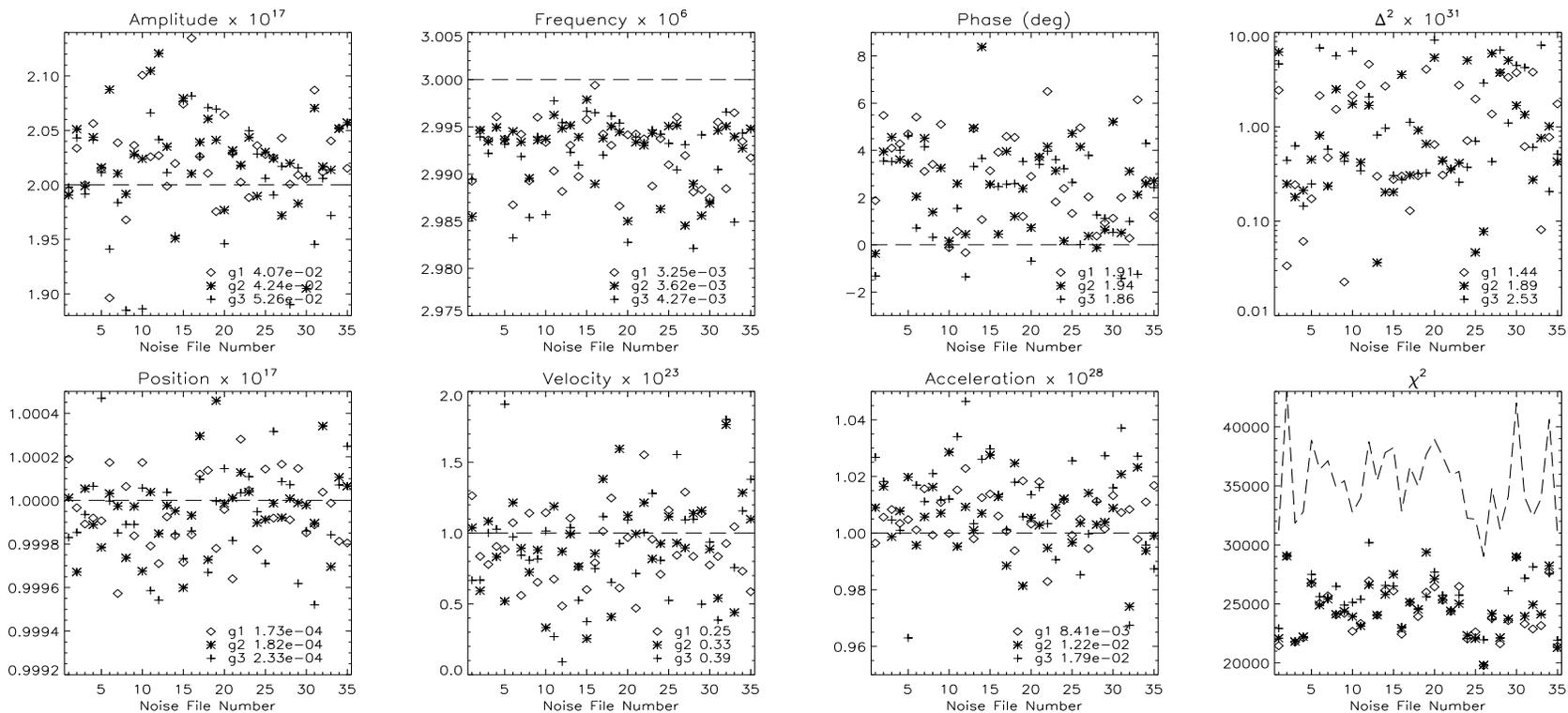, height=9in, width=4.in, angle=90}
\caption{\label{ge3fig} Search algorithm results for 
the 35 realizations of the LISA data stream
with a SNR=5 $3~\uHz$ signal and a disturbance with an 
interruption length of 17~minutes.  The 
dashed lines represent the input parameters.  
In the case of $\chi^2$, the dashed
line is the value of $\chi^2$ in the original LISA data stream realization.
As in the other figures, the standard deviations are 
printed at the bottom of each plot.}
\end{sidewaysfigure}

Comparing the disturbance values in this table with those in Table~\ref{gapOnly}
we see that the derived positions and accelerations are very close, 
but the velocities are different.
Comparing the results for all 35 noise realizations we see that
in Figure~\ref{gapOnlyFig} (which contains a disturbance by no signal)
there is a systematic low in the velocity,
whereas in Figure~\ref{ge3fig} (which contains both a disturbance and
a signal) the systematic low has been removed.
The Fourier transform of a linear slope is a $1/f$ power spectrum.
A monochromatic sine wave will have $1/f$ wings.  Therefore it is 
not unreasonable for there to be a correlation between the frequency of the
sinusoid (which determines the amplitude of the wings) and the velocity.
This correlation has caused the derived velocities to be different when
there is a sine wave present.  

\begin{table}[t]
\caption[1000 Second Data Results]{\label{ge3}
Search algorithm results for the three disturbances examined,
and with a low frequency signal.
$A$ is in multiples of $10^{-17}$, $f^* = (f - 3\uHz) / \mathrm{nHz}$,
$p^* = (p - 10^{-17}) \times 10^{20}$, 
$u$ is in multiples of $10^{-23}$, $a$ is in multiples of $10^{-28}$,
and $\Delta^2$ is in multiples of $10^{-31}$.}
\begin{indented}\lineup
\item[]\begin{tabular}{@{}|c||rrr|rrr||cc|}
\hline
 & \multicolumn{1}{c}{$A$} & \multicolumn{1}{c}{$f^*$} & \multicolumn{1}{c|}{$\phi$} & 
   \multicolumn{1}{c}{$p^*$} & \multicolumn{1}{c}{\0$u$} & \multicolumn{1}{c||}{$a$} & 
   $\chi^2$ & $\Delta^2$ \\
\hline \hline
\input{ge3.out}
\hline
\end{tabular}\end{indented}\end{table}

Figure~\ref{ge3fig} contains results for all 35 noise realizations.  From these
results we again compute standard deviations and use these as estimates
of the errors in each of our derived parameters.  The errors in the signal
parameters are roughly 
$(2.3\%, 0.12\%, 0.53\%)$. 
The errors in the disturbance parameters are roughly
$[0.02\%, 32\%, 1.3\%]$.
With a disturbance present our algorithm has performed worse at obtaining the
true values of the signal parameters.  
In addition, the spread in derived disturbance parameters has also
increased.
However, the derived errors from the 35 realizations 
imply that the input parameters are just within our solution.  
This does not mean that our algorithm did a good job at
locating the signal, rather that the disturbance did not significantly
push the solution away from the input.

\subsection{Higher Frequency Signal}

Disturbances should have a stronger effect on low frequency sources than high
frequency sources.  At higher frequencies, data disturbances should be less
noticeable since there are many more cycles present in the data set.  Fitting
for the disturbance parameters with only a high frequency signal present
should be much like fitting for the parameters with no signal present at all.

\begin{table}[!ht]
\caption[1000 Second Data Results]{\label{he3}
Search algorithm results for the three disturbances examined,
and with a high frequency signal.
$A$ is in multiples of $10^{-21}$, $f^* = (f - 100\uHz) / \mathrm{nHz}$,
$p^* = (p - 10^{-17}) \times 10^{20}$, 
$u$ is in multiples of $10^{-23}$, $a$ is in multiples of $10^{-28}$,
and $\Delta^2$ is in multiples of $10^{-39}$.}
\begin{indented}\lineup
\item[]\begin{tabular}{@{}|c||rrr|rrr||cc|}
\hline
 & \multicolumn{1}{c}{$A$} & \multicolumn{1}{c}{$f^*$} & \multicolumn{1}{c|}{$\phi$} & 
   \multicolumn{1}{c}{$p^*$} & \multicolumn{1}{c}{\0$u$} & \multicolumn{1}{c||}{$a$} & 
   $\chi^2$ & $\Delta^2$ \\
\hline \hline
\input{he3.out}
\hline
\end{tabular}\end{indented}\end{table}

We choose to look at a 100~$\uHz$ signal for our higher frequency case.  
A signal-to-noise ratio of 5 implies an rms strain amplitude of 
$1.6 \times 10^{-21}$ for our realization.
Our input parameters are now
$(A, f, \phi)_\mathrm{input}~=~(1.6\times10^{-21}, 100\uHz, 0^\circ)$.
With no disturbance, our search algorithm found the following parameters:
$(1.682110\times10^{-21},~99.98995\uHz,~10.60217^\circ)$
with $\chi^2 = 30692$.
Table~\ref{he3} contains results for this high frequency signal in the 
presence of a disturbance with an interruption of 17 minutes in length.
Comparing with Table~\ref{gapOnly} we see that 
the derived disturbance parameters are
almost identical.  The frequency of the signal has been found to within 10~nHz,
or 0.015\%, of the input frequency.  The disturbance has had little effect
on the ability of our algorithm to determine 
the high frequency signal parameters.

\section{Disturbance Limitations on Operations}\label{limitations}

As we have seen in the previous section, isolated disturbances do not
significantly affect our ability to identify and remove sources
from the data stream, given enough data before and after
the disturbance location.  The following question can now be posed:
How does the frequency of disturbances affect our search algorithm?  
For this analysis we examined only the lower frequency signal case
since the identification of higher frequency signals is not perturbed
much in the presence of a disturbance.  

\subsection{Frequency}

Each disturbance contains three parameters to be fitted for.
As mentioned earlier, the accuracy to which you can determine the parameters
depends on the number of sample points available.  Within a 38-day period,
the inclusion of multiple disturbances effectively reduces the amount of 
information
available to each disturbance.  We can simulate the effect of having
multiple disturbances by placing one disturbance near the end of the data 
stream.
This assumes that all previous disturbances have been fit and removed.
Table~\ref{lastgap} shows our results for one disturbance, with $L=1000$ 
seconds, 
at various distances from the end of the data stream.
With a small number of points after the data interruption it is hard to 
determine
the acceleration parameter.  This implies that the assumption that we removed
all other disturbances exactly is wrong---we probably would not have been
able to determine the acceleration well.  
Therefore we must fit for all of the disturbances
at once since the acceleration parameter will strongly affect the data at
large times.

\begin{table}
\caption[One Disturbance]{\label{lastgap}
Search results for a low frequency signal with one disturbance.
The disturbance start time is $T$ before the end of the data stream,
and the length $L$ is 1000 seconds.  The last three data sets are $g3$,
$g2$, and $g1$ from Table~\ref{ge3}.
$A$ is in multiples of $10^{-17}$, $f^* = (f - 3\uHz)/\mathrm{nHz}$, 
$p^* = (p - 10^{-17}) \times 10^{20}$,
$u$ is in multiples of $10^{-23}$, $a$ is in multiples of $10^{-28}$, 
and $\Delta^2$ is in multiples of $10^{-31}$.}
\begin{indented}\lineup
\item[]\begin{tabular}{@{}|c||crr|rrr||cc|}
\hline
$T$ & $A$ & \multicolumn{1}{c}{$f^*$} & \multicolumn{1}{c|}{$\phi$} & 
   \multicolumn{1}{c}{\0\0$p^*$} & \multicolumn{1}{c}{\0$u$} & 
   \multicolumn{1}{c||}{$a$} & $\chi^2$ & $\Delta^2$ \\
\hline \hline
\input{loc.out}
\end{tabular}\end{indented}\end{table}

As a preliminary investigation we examined several cases 
with multiple disturbances.
We simulated data streams with the maximal number of disturbances
possible all equally spaced by $T$, and of length $L=1000$ seconds.
Our results of this analysis are in Table~\ref{multigaps}.  Note
that our search algorithm was not designed to handle more 
than 20 disturbances, therefore instead of having 31 equally
spaced disturbances in the $T=1.2$ days data set and 37 disturbances
in the $T=1$ case, we only have 20.

It appears that our algorithm has trouble with disturbance frequencies of
more than one every 3.5 days.  The algorithm not only has trouble 
fitting for the frequency and phase, 
but the velocities and accelerations for most of the the disturbances are 
much worse---this is evident by the large values of $\chi^2$.  
Some refinement of the search algorithm
for multiple gaps may help to fit the velocities and accelerations 
better for a high frequency of disturbances.

\begin{table}
\caption[Multiple Disturbances]{\label{multigaps}
Search algorithm results in the presence of multiple disturbances.
\# is the number of disturbances present in the data stream, separated
in time by $T$, and all of length $L=1000$ seconds.
Notice the degradation in the derived signal parameters and the value of
$\chi^2$ for disturbance frequencies more than one every 3.5 days.
$A$ is in multiples of $10^{-17}$, $f^* = (f - 3\uHz)/\mathrm{nHz}$, 
and $\Delta^2$ is in multiples of $10^{-31}$.}
\begin{indented}\lineup
\item[]\begin{tabular}{@{}|c||c||crr||rr|}
\hline
$T$ & \# & $A$ & \multicolumn{1}{c}{$f^*$} & \multicolumn{1}{c||}{$\phi$} & 
   \multicolumn{1}{c}{$\chi^2$} & \multicolumn{1}{c|}{$\Delta^2$} \\
\hline \hline
\input{locAll.out}
\end{tabular}\end{indented}\end{table}

\section{Conclusions}

We have presented an algorithm which identifies monochromatic signals 
in the presence of data disturbances in a LISA-like data stream.
The results here are encouraging on the ability 
of removing the disturbances without affecting the science results.  
However, we found complications with data sets containing 
disturbances which are more frequent than about one every 3.5 days.
There are a number of simplifications in our data stream, for instance
the use of monochromatic signals and instantaneous drop out and recovery
on the edges of interruptions.
As mentioned in \S\ref{nosig} longer data sets will allow all parameters
to be fit more accurately.  Therefore the results from our short 38 day 
long data set analysis should be taken as a guide for longer data sets.

\ack

Support for this work has been provided under NGT5-50451 and S-73625-G.  
We thank Peter L. Bender and Andrew J. S. Hamilton
for the many discussions on this project 
and for their numerous comments on this manuscript.
Computational support has been graciously provided by the
W. M. Keck Foundation.

\end{document}